\documentclass{article}
\begin{document}

\author{J.J. Rosales$^{a} \footnote{E-mail: rosales@salamanca.ugto.mx}$ \ and V.I. Tkach$^b$\\
$^a$Departamento de Ingenier\'ia El\'ectrica \\
Divisi\'on de Ingenier\'ias Campus Irapuato-Salamanca\\
Universidad de Guanajuato\\
Carretera Salamanca-Valle de Santiago, km. 3.5 + 1.8 km\\
Comunidad de Palo Blanco, Salamanca Guanajuato. M\'exico\\
\\
$^b$ Department of Physics and Astronomy\\
Northwestern University\\
Evanston, IL 60208-3112, USA}
\title{Supersymmetric Cosmological FRW Model and Dark Energy}
\maketitle

In this work we consider a flat cosmological model with a set of
fluids in the framework of supersymmetric cosmology. The obtained
supersymmetric algebra allowed us to take quantum solutions. It is
shown that only in the case of a cosmological constant we have a
condition between the density of dark energy $\rho_\Lambda$ and
density energy of matter $\rho_M$, $\rho_\Lambda
> 2\rho_M$.
\\

\\

Minisuperspace models are useful toy models for canonical quantum
gravity, because they capture many of the essential features of
general relativity and are at the same time free of technical
difficulties associated with the presence of an infinite number of
degrees of freedom. As it is well known, the equation that governs
the quantum behavior of these models is the Wheeler-DeWitt
equation, which results in a quadratic Hamiltonian leading to an
equation of the Klein-Gordon type. Introduction of supersymmetric
minisuperspace models in which the Grassmann variables are not
identified as the supersymmetric partners of the cosmological
bosonic variables has led to the definition and study of linear
"square root" equations defining the quantum evolution of the
universe $[1-11]$.


Recently, we have used the superfield formulation to investigate
supersymmetric cosmological models $[12-15]$. In previous works
\cite{14,16} it was shown that the spatially homogeneous part of
the fields in the supergravity theory preserves the invariance
under the local time $n = 2$ supersymmetry. This supersymmetry is
a subgroup of the four-dimensional spacetime supersymmetry of the
supergravity theory. This local supersymmetry procedure has the
advantage that, by defining the superfields on superspace, all the
component fields in a supermultiplet can be manipulated
simultaneously in a manner that automatically preserves
supersymmetry. Besides, the Grassmann variables are obtained in a
clear manner as the supersymmetric partners of the cosmological
bosonic variables. At the quantum level the Grassmann variables
are elements of the Clifford algebra. Using superfield formulation
the canonical quantization procedure for a closed FRW cosmological
model, filled with pressureless matter (dust) content and the
corresponding superpartner, was reported \cite{16}.

In the present work we have constructed the $n = 2$ supersymmetric
action for the spatially homogeneous isotropic flat, ($k = 0$)
Friedmann-Robertson-Walker including a mixture of fluids with a
constant equation of state parameters $\gamma_i$, $p_i =
\gamma_i\rho_i$.

Let us start with the action \cite{17,18}
\begin{equation}
S = \int\Big[ -\frac{R}{2N \tilde G}\Big(\frac{dR}{dt}\Big)^2 +
\frac{N \Lambda}{6 \tilde G} R^3 + N \Big(\sum_i
M^{1/2}_{\gamma_i}R^{-{3\gamma_i}/2}\Big)^2\Big] dt, \label{2}
\end{equation}
with $ \tilde G = \frac{8 \pi G}{6}$, where $G$ is the Newtonian
gravitational constant and $\Lambda$ is the cosmological constant;
$N(t), R(t)$ are the lapse function and the scale factor,
respectively; $M_{\gamma_i}$ is the mass by unit $({\rm
length})^{3\gamma_i - 1}$. Summation over $i$ includes all types
of fluids. In this work we have used units in which $c=\hbar = 1$.

The action (\ref{2}) is invariant under the time reparametrization
$t^{\prime} \to t + a(t), \label{3}$
if the transformations of $R(t)$ and $N(t)$ are defined as
$\delta R = a {\dot R}$, and $\delta N = (aN)^{.}\label{4}$.
The variation with respect to $R(t)$ and $N(t)$ leads to the
classical equation for the scale factor $R(t)$ and the constraint,
which generates the local reparametrization of $R(t)$ and $N(t)$.
This constraint leads to the Wheeler-DeWitt equation in quantum
cosmology.

In order to obtain the corresponding supersymmetric action for
(\ref{2}), we follow the superfield approach. Thus, we extend the
transformation of time reparametrization to the $n= 2$ local
supersymmetry of time $(t, \eta, \bar\eta)$. Then, we have the
following local supersymmetric transformation
\begin{eqnarray}
\delta{t}  &  = & a(t) + \frac{i}{2}[\eta{\bar\beta^{\prime}}(t) +
\bar\eta\beta^{\prime}(t)],\nonumber\\
\delta\eta &  = & \frac{1}{2}\beta^{\prime}(t) + \frac{1}{2} [\dot
a(t) + ib(t)]\eta + \frac{i}{2} \dot\beta^{\prime}(t) \eta\bar\eta
,\label{5}\\
\delta{\bar\eta}  &  = & \frac{1}{2}\bar\beta^{\prime}(t) +
\frac{1}{2}[\dot a(t) - ib(t)]\bar\eta-
\frac{i}{2}\dot{\bar\beta}^{\prime}(t) \eta\bar\eta,\nonumber
\end{eqnarray}
where $\eta$ is a complex Grassmann coordinate, $\beta^{\prime}(t)
= N^{-1/2}\beta(t)$ is the Grassmann complex parameter of the
local ``small'' $n=2$ supersymmetry (SUSY) transformation, and
$b(t)$ is the parameter of local $U(1)$ rotations of the complex
$\eta$. The superfield generalization of the action (\ref{2}),
invariant under supersymmetric transformation (\ref{5}) has the
form
\begin{eqnarray}
S_{susy} &=& \int \Big[ -\frac{1}{2 \tilde G} {I\!\!N}^{-1}
{I\!\!R}D_{\bar\eta}{I\!\!R} D_\eta{I\!\!R} + \frac{
\Lambda^{1/2}}{3\sqrt{3}\tilde G}{I\!\!R}^3
-\nonumber\\
&-&\frac{2 \sqrt{2}}{\tilde G^{1/2}} \sum_i
\frac{M^{1/2}_{\gamma_i}}{(3 - 3\gamma_i)}
{I\!\!R}^{\frac{3-3\gamma_i}{2}}\Big]d{\eta} d{\bar\eta}dt,
\label{6}
\end{eqnarray}
where
\begin{equation}
D_{\eta} = \frac{\partial }{\partial \eta} + i\bar\eta
\frac{\partial }{\partial t}, \qquad D_{\bar\eta} =
-\frac{\partial }{\partial \bar\eta} - i\eta \frac{\partial
}{\partial t}, \label{7}
\end{equation}
are the supercovariant derivatives of the global "small"
supersymmetry of the generalized parameter corresponding to $t$.
The local supercovariant derivatives have the form ${\tilde
D}_{\eta} = {I\!\!N}^{-1/2} D_\eta$, ${\tilde D}_{\bar\eta} =
{I\!\!N}^{-1/2} D_{\bar\eta}$, and ${I\!\!R}(t,\eta, \bar\eta),
{I\!\!N}(t,\eta, \bar\eta)$ are superfields. The Taylor series
expansion for the superfields ${I\!\!N}(t,\eta,\bar\eta)$ and
${I\!\!R}(t,\eta,\bar\eta)$ is the following
\begin{eqnarray}
{I\!\! N}(t,\eta,\bar\eta)&=& N(t) + i\eta\bar\psi^{\prime}(t) +
i\bar
\eta\psi^{\prime}(t) + V^{\prime}(t)\eta\bar\eta,\label{8}\\
{I\!\! R}(t,\eta,\bar\eta) &=& R(t) +
i\eta\bar\lambda^{\prime}(t)+ i\bar\eta\lambda^{\prime}(t) +
B^{\prime}(t) \eta\bar\eta.\label{9}
\end{eqnarray}
In these expressions we have introduced the redefinitions
$\psi^{\prime}(t) = N^{1/2}\psi(t)$, $V^{\prime} = N(t)V(t) +
\bar\psi(t) \psi(t)$, $\lambda^{\prime} = \frac{\tilde
G^{1/2}N^{1/2}}{R^{1/2}} \lambda$ and $B^{\prime} = \tilde
G^{1/2}NB + \frac{\tilde G^{1/2}}{2R^{1/2}}(\bar\psi \lambda -
\psi \bar\lambda)$. The components of the superfield ${I\!\!N}(t,
\eta, \bar\eta)$ are gauge fields of the one-dimensional $n=2$
extended supergravity. $N(t)$ is the einbein, $\psi(t),
\bar\psi(t)$ are the complex gravitino fields, and $V(t)$ is the
$U(1)$ gauge field. The component $B(t)$ in (\ref{9}) is an
auxiliary degree of freedom (non-dynamical variable), and
$\lambda, \bar\lambda$ are the fermion partners of the scale
factor $R(t)$.

After integration over the Grassmann coordinates $\eta, \bar\eta$
and eliminating the auxiliary variable $B$, by means of their
equation of motion, the action (\ref{6}) acquires its component
form
\begin{eqnarray}
S_{susy} &=& \int \left\{- \frac{R (DR)^2}{2N \tilde G} + \frac{N
\Lambda R^3}{6 \tilde G} + N \Big(\sum_i M^{1/2}_{\gamma_i}
R^{-3/2\gamma_i}\Big)^2 -
\right.\nonumber\\
&& \left. - \frac{\sqrt{2}\Lambda^{1/2}}{\sqrt{3} \tilde G^{1/2}}
\sum_iM^{1/2}_{\gamma_i} R^{\frac{3 - 3 \gamma_i}{2}} +
\frac{i}{2}(\bar\lambda D\lambda - D\bar\lambda \lambda) +
\frac{\sqrt{3}}{2}\Lambda^{1/2}N \bar\lambda \lambda
\right.\nonumber\\
&& \left. + \frac{N\sqrt{2}}{\tilde G^{1/2}} \sum_i (-1 +
6\gamma_i) M^{1/2}_{\gamma_i} R^{\frac{-3 - 3
\gamma_i}{2}}\bar\lambda \lambda\right. \label{13}
\\
&& \left. + \frac{\Lambda^{1/2}}{2\sqrt{3}\tilde G^{1/2}}
R^{3/2}(\bar\psi \lambda - \psi \bar\lambda) - \frac{\sqrt{2}}{2}
\sum_iM^{1/2}_{\gamma_i}R^{-\frac{3\gamma_i}{2}}(\bar\psi \lambda
- \psi \bar\lambda) \right\}dt,  \nonumber
\end{eqnarray}
with $DR = {\dot R} - \frac{i\tilde G^{1/2}}{2R^{1/2}}(\psi
\bar\lambda + \bar\psi \lambda)$ and $D\lambda = {\dot \lambda} -
\frac{1}{2} V\lambda$, $D{\bar\lambda} = {\dot {\bar\lambda}} +
\frac{1}{2} V\bar\lambda$. Proceeding with canonical quantization,
the classical canonical Hamiltonian is calculated in the usual way
for systems with constraints. It has the form
\begin{equation}
H_c = NH + \frac{1}{2} \bar\psi S - \frac{1}{2}\psi {\bar S} +
\frac{1}{2}VF, \label{14}
\end{equation}
where $H$ is the Hamiltonian of the system, $S$ and ${\bar S}$ are
the supercharges and $F$ is the $U(1)$ rotation generator. The
canonical Hamiltonian form (\ref{14}) explains the fact that $N,
\psi, \bar\psi$ and $V$ are Lagrangian multipliers, which only
enforce the first-class constraints $H = 0, S = 0, {\bar S } = 0$
and $F = 0$, and express the invariance under the $n = 2$
supersymmetric transformations. The first-class constraints may be
obtained from the action (\ref{13}) rewriting it in first order
form varying $N(t),\psi(t)$,$\bar\psi(t)$ and $V(t)$,
respectively. The first-class constraints are
\begin{eqnarray}
H &=& -\frac{\tilde G}{2R} \pi^2_R - \frac{\Lambda R^3}{6 \tilde
G} - \Big(\sum_i M^{1/2}_{\gamma_i}R^{-3/2\gamma_i}\Big)^2 +
\frac{\sqrt{2}\Lambda^{1/2}}{\sqrt{3} \tilde G^{1/2}}
\sum_iM_{\gamma_i}^{1/2} R^{\frac{3 - 3\gamma_i}{2}}\nonumber\\
&-& \frac{\sqrt{3}}{2}\Lambda^{1/2} \bar\lambda \lambda -
\frac{1}{\sqrt{2} \tilde G^{1/2}}\sum_i (6\gamma_i - 1)
M_{\gamma_i}^{1/2}R^{\frac{-3 - 3 \gamma_i}{2}}\bar\lambda \lambda
,\label{15}\\
S&=& \Big(\frac{i\tilde G^{1/2}}{R^{1/2}}\pi_R -
\frac{ \Lambda^{1/2} R^{3/2}}{\sqrt{3}\tilde G^{1/2}} +
\sqrt{2}\sum_i M_{\gamma_i}^{1/2} R^{-\frac{3\gamma_i}{2}}\Big)
\lambda, \label{16} \\
{\bar S}&=& \Big(-\frac{i\tilde G^{1/2}}{R^{1/2}}\pi_R - \frac{
\Lambda^{1/2} R^{3/2}}{\sqrt{3}\tilde G^{1/2}} + \sqrt{2}\sum_i
M_{\gamma_i}^{1/2} R^{-\frac{3\gamma_i}{2}}\Big)\bar\lambda, \label{17}\\
F&=& - \bar\lambda \lambda, \label{18}
\end{eqnarray}
where $\pi_R = - \frac{R}{{\tilde G}N} {\dot R} +
\frac{iR^{1/2}}{2N \tilde G^{1/2}}(\bar\psi \lambda + \psi
\bar\lambda)$ is the canonical momentum associated to $R$ with
Poisson brackets
\begin{equation}
\lbrace R, \pi_R \rbrace = 1. \label{19}
\end{equation}
As usually with Grassmann variables, we have second-class
constraints, which can be eliminated by Dirac procedure. As a
result, we only have the following non-zero Dirac brackets
\begin{equation}
\lbrace \lambda, \bar\lambda\rbrace = i. \label{19a}
\end{equation}
With respect to these brackets the super-algebra for the
generators $H, S, {\bar S}$ and $F$ becomes
\begin{equation}
\lbrace S, {\bar S}\rbrace = - 2iH,\quad \lbrace S, H \rbrace =
\lbrace {\bar S}, H \rbrace = 0, \quad \lbrace F, S\rbrace = iS.
\label{20}
\end{equation}
In a quantum theory the brackets (\ref{19}) and (\ref{19a}) must
be replaced by commutator and anticommutator; they can be
considered as generators of the Clifford algebra
\begin{equation}
[R, \pi_R]= i \, \qquad {\rm with} \quad \ \pi_R = -i
\frac{\partial}{\partial R} \label{21},\qquad \lbrace \lambda,
\bar\lambda \rbrace = - 1.
\end{equation}
As we can see from the Hamiltonian (\ref{15}), the energy of
scalar factor $R$ is negative. This is reflected in the fact that
the anticommutator value (\ref{21}) of superpartner $\lambda$ and
$\bar\lambda$ of the scalar factor is negative. This
anticommutator relation may be satisfied under condition
\begin{eqnarray}
\bar\lambda &=& \xi^{-1} \lambda^{\dagger}\xi = -
\lambda^{\dagger},\qquad \lbrace \lambda, \lambda^{\dagger}\rbrace
= 1,\nonumber\\
\lambda^{\dagger} \xi &=&\xi \lambda^{\dagger} \qquad {\rm and}
\qquad \xi^{\dagger} = \xi. \label{20a}
\end{eqnarray}
So, for the supercharge operator $\bar S$ we have $\bar S =
\xi^{-1} S^{\dagger} \xi$. The quantum generators $H,S,\bar S$ and
$F$ form a closed super-algebra
\begin{equation}
\lbrace S, {\bar S}\rbrace = 2H,\quad [ S, H ]= [{\bar S}, H ]= 0,
\quad [ F, S] = -S,\quad S^2 = 0. \label{20b}
\end{equation}
In the case of standard supersymmetric quantum mechanics, we could
have $\bar\lambda = \lambda^{\dagger}$, $\bar S = S^{\dagger}$ and
the Hamiltonian would be positive. We can see, that the
anticommutator of supercharges $S$ and their conjugated ${\bar S}$
under our conjugated operation (\ref{20a}) has the form
$\overline{ \lbrace S,{\bar S}\rbrace} = \lbrace S, {\bar
S}\rbrace$ and the Hamiltonian operator is self-conjugated $\bar H
= H$ under the operation $\bar H = \xi^{-1} H^{\dagger}\xi$. We
can fulfill them on the Fock space representation with
$\bar\lambda$ as a creation and $\lambda$ as annihilation vacuum
operators on the Fock space, then the general quantum state can be
written as vectors depending on $R$ in the corresponding Fock
space \cite{14}.

We can choose the matrix representation for the fermionic
para\-meters $\lambda, \bar\lambda$ and $ \xi$ as $\lambda =
\sigma_-, \quad \bar \lambda = - \sigma_+, \quad \xi= \sigma_3$,
with $\sigma_\pm = \frac{1}{2}(\sigma_1 \pm i\sigma_2)$, where
$\sigma_1$, $\sigma_2$ and $\sigma_3$ are the Pauli matrices. The
supercharges $S$, $\bar S$ have the following structures
\begin{equation}
S = A\lambda, \qquad\qquad S^{\dag} = A^{\dag} \lambda^{\dag},
\label{26}
\end{equation}
where
\begin{equation}
A = i\tilde G^{1/2} R^{-1/2} \pi_R - \frac{\Lambda^{1/2}
R^{3/2}}{\sqrt{3}\tilde G^{1/2}} + \sqrt{2} \sum_i
M_{\gamma_i}^{1/2}R^{-\frac{3\gamma_i}{2}}.\label{27}
\end{equation}
An ambiguity exist in the factor ordering of these operators, such
ambiguities always arise when the operator expression contains the
product of non-commuting operators $R$ and $\pi_R$, as in our
case. It is then necessary to find some criteria to know which
factor ordering should be selected. We propose the following; to
integrate with measure $R^{1/2}dR$ in the inner product of two
states \cite{15,16}. In this measure the conjugate momentum
$\pi_R$ is non-Hermitian with $\pi^{\dagger}_R = R^{-1/2}\pi_R
R^{1/2}$. However, the combination $(R^{-1/2}\pi_R)^{\dagger} =
R^{-1/2}\pi_R $ is Hermitian one, and
$(R^{-1/2}\pi_RR^{-1/2}\pi_R)^{\dagger} = R^{-1/2}\pi_R
R^{-1/2}\pi_R$ is Hermitian too.

In the quantum theory, the first-class constraints become
conditions on the wave function $\Psi(R)$. Furthermore, any
physical state must be satisfy the quantum constraints
\begin{equation}
H \Psi(R) = 0, \quad S\Psi(R) = 0,\quad {\bar S}\Psi(R) = 0, \quad
F\Psi(R) = 0, \label{29}
\end{equation}
where the first equation is the Wheeler-DeWitt equation for the
minisuperspace model. The eigenstates of the Hamiltonian have two
components in the matrix representation
\begin{equation}
\Psi = \pmatrix{\Psi_1\cr \Psi_2 \cr}. \label{30}
\end{equation}
However, the supersymmetric physical states are obtained applying
the supercharges operators $S\Psi = 0, {\bar S}\Psi = 0$. Using
the algebra given by (\ref{20b}), these are rewritten in the
following form
\begin{equation}
(\lambda {\bar S} - \bar\lambda S)\Psi = 0.\label{31}
\end{equation}
Using the matrix representation for $\lambda$ and $\bar\lambda$ we
obtain the following differential equations for $\Psi_1(R)$ and
$\Psi_2(R)$ components
\begin{eqnarray}
\Big({\tilde G^{1/2}} R^{-1/2} \frac{\partial }{\partial R} -
\frac{\Lambda^{1/2} R^{3/2}}{\sqrt{3} \tilde G^{1/2}} +
\sqrt{2}\sum_i M_{\gamma_i}^{1/2}R^{-\frac{3\gamma_i}{2}}
\Big) \Psi_1(R) = 0,\label{32}\\
\Big(\tilde G^{1/2} R^{-1/2} \frac{\partial }{\partial R} +
\frac{\Lambda^{1/2} R^{3/2}}{\sqrt{3} \tilde G^{1/2}} - \sqrt{2}
\sum_i M_{\gamma_i}^{1/2}R^{-\frac{3\gamma_i}{2}} \Big) \Psi_2(R)
= 0.\label{33}
\end{eqnarray}
Solving these equations and using the relation $M_{\gamma_i} =
\frac{1}{2} \rho_{\gamma_i} R^{3(\gamma_i + 1)}$, we have the
following solutions
\begin{eqnarray}
\Psi_1(R) = C_1 \exp{\Big[ \frac{1}{\sqrt{6\pi}}\Big(
\frac{\rho_\Lambda}{\rho_{pl}}\Big)^{1/2} \Big(
\frac{R}{l_{pl}}\Big)^3 -
\frac{\sqrt{18}}{\sqrt{6\pi}}\frac{1}{\rho_{pl}^{1/2}}\Big(
\frac{R}{l_{pl}}}\Big)^3 \sum_i \frac{\rho_{\gamma_i}^{1/2}}{(3 -
3\gamma_i)} \Big], \label{35}
\end{eqnarray}
\begin{eqnarray}
\Psi_2(R) = C_2 \exp{\Big[ -\frac{1}{\sqrt{6\pi}}\Big(
\frac{\rho_\Lambda}{\rho_{pl}}\Big)^{1/2} \Big(
\frac{R}{l_{pl}}\Big)^3 +
\frac{\sqrt{18}}{\sqrt{6\pi}}\frac{1}{\rho_{pl}^{1/2}}\Big(
\frac{R}{l_{pl}}}\Big)^3 \sum_i \frac{\rho_{\gamma_i}^{1/2}}{(3 -
3\gamma_i)}\Big], \label{36}
\end{eqnarray}
where $\rho_{pl} = G^{-2}$ is the Planck density and $l_{pl} =
G^{1/2}$ is the Planck length. We can see, that the function
$\Psi_1$ in (\ref{35}) has good behavior when $R \to \infty$ under
the condition $\rho_\Lambda < 18\Big(\sum_i
\frac{\rho_{\gamma_i}^{1/2}}{(3 - 3\gamma_i)}\Big)^2$, while
$\Psi_2$ does not. On the other hand, the wave function $\Psi_2$
in (\ref{36}) has good behavior when $R \to \infty$ under the
condition
\begin{equation}
\rho_\Lambda > 18\Big(\sum_i \frac{\rho_{\gamma_i}^{1/2}}{(3 -
3\gamma_i)}\Big)^2 \label{37},
\end{equation}
because the main contribution comes from the first term of the
exponent, while $\Psi_1$ does not have good behavior. However, the
wave function $\Psi$ in the state with zero energy; $\Psi^{T} =
(0, \Psi_2)$ is normalizable in the measure $R^{1/2} dR$ under the
condition (\ref{37}), such as $H\Psi = S\Psi = \bar S \Psi = F\Psi
= 0$ with $F = \sigma_{+} \sigma_{-}$.

If $H\Psi = E\Psi$, the eigenstate $\Psi_1$ (\ref{35}) of the
quantum Hamiltonian (\ref{15}) for $E = 0$ is non-normalizable.
But for non-zero eigenvalues of the Hamiltonian (\ref{15}) it is
known that there exist two normalizable components (wave
functions).

The condition ({\ref{37}) does not contradict the astrophysical
observation at $\rho_{\Lambda} \approx (2\sim 3)\rho_M$, due to
the fact that the dust-like matter introduces the main
contribution to the total energy density of matter $\rho_M =
\sum_{i}\rho_{\gamma_i}\approx \rho_{\gamma  = 0}$. We have from
(\ref{37}) the pressureless fluid $(\gamma = 0)$ contents barionic
and cold dark matter $(\rho_\Lambda > 2\rho_M)$.

On the other hand, according to recent astrophysical data, our
universe is dominated by a mysterious form of dark energy
\cite{19}, which counts up to about $70$ per cent of the total
energy density. As a result, the universe expansion is
accelerating \cite{20,21}. Vacuum energy density $\rho_\Lambda =
\frac{\Lambda}{8\pi G}$ is a concrete example of dark energy
\cite{22}.

The recent cosmological data give us the following range for the
dark energy state parameter $-1.14< \gamma < - 0.88$. However, in
the literature we can find different theoretical models
\cite{23,24} for the dark energy with state parameter $\gamma
> - 1$ and $\gamma < - 1$. In the case $\Lambda = 0$ we see from (\ref{35}), that
the wave function $\Psi$; $\Psi^{T} = (\Psi_1, 0)$ is
normalizable. Moreover, if we assume that the universe may enter
into phantom phase $(\gamma < - 1)$ or quintessence phase $(\gamma
> - 1)$ we don't have conditions between density of dark energy
$\rho_{ph}(\gamma < -1)$, $\rho_{q}(-1< \gamma)$ and density of
energy matter $\rho_M$.

\end{document}